\def \cm{~\rm{cm}}
\def \s{~\rm{s}}
\def \ms{~\rm{ms}}
\def \km{~\rm{km}}

\def \K{~\rm{K}}
\def \g{~\rm{g}}

\def \erg{~\rm{erg}}

\documentclass[12pt,preprint]{aastex}
\usepackage{natbib}
\usepackage{amsmath}                
\usepackage{amsfonts}               
\usepackage{amssymb}                
\usepackage{epsfig}                 

\shortauthors{Soker}

\begin{document}

\title{EXPLODING CORE-COLLAPSE SUPERNOVAE WITH JITTERING JETS}

\author{Oded Papish and Noam Soker\altaffilmark{1}}

\altaffiltext{1}{Dept. of Physics, Technion, Haifa 32000, Israel;
papish@techunix.technion.ac.il; soker@physics.technion.ac.il.}

\begin{abstract}
We argue that jittering jets, i.e., jets that have their launching direction rapidly change,
launched by the newly formed neutron star in a core collapse supernova can explode the star.
We show that under a wide range of parameters the fast narrow jets deposit their energy
inside the star via shock waves, and form two hot bubbles, that eventually merge, accelerate the rest
of the star and lead to the explosion.
To prevent the jets from penetrating through the collapsing stellar core and escape with their energy,
instead of forming the hot bubbles, the jets should be prevented from drilling a hole through the star.
This condition can be met if the jets' axis rapidly changes its direction.
This process of depositing jets' energy into the ambient medium is termed the
{\it penetrating jet feedback mechanism.}
The feedback exists in that the neutron star (or a black hole) at the center of the core collapse supernova shuts
off its own growth by exploding the star.
The jets deposit their energy at a distance of $\sim 1000 \km$ from the center and expel the mass above that radius.
In our model, the material near the stalled shock at several hundreds kilometers from the center
is not expelled, but it is rather accreted and feed the accretion disk that blows the jets.
The neutrinos might influence the accretion flow, but in the proposed model their role in exploding the star is small.
\end{abstract}


\section{INTRODUCTION}
\label{sec:intro}

Jets can play a key role in exploding core collapse supernovae (CCSNe; e.g., Khokhlov et al. 1999;
MacFadyen et al. 2001; Maeda \& Nomoto 2003; Woosley \& Janka 2005; Couch et al. 2009).
Khokhlov et al. (1999) injected a jet with a radius of $r_{j0}=1200 \km$ at a distance of
$R_{\rm in}=3820\km$ from the center.
The mass in the two jets was $\sim 0.1 M_\odot$, and their speed $\sim 0.1 c$.
Practically they injected \emph{slow massive wide} (SMW) jets to explode the star.
As the jets launched by the newly formed neutron star (NS) are likely to be narrow and fast
(a velocity of $>0.1 c$), the formation of the jets launched by Khokhlov et al. (1999)
should be explained.

Couch et al. (2009) found that to match observations their jets must start with
a large fraction of thermal energy (their model v1m12).
The total mass in the jets in their model v1m12 was $0.12 M_\odot$, and the maximum velocity
$10^4 \km \s^{-1}$.
As evident from their fig. 7, such jets practically start as wide jets.
Namely, their initial conditions for their model v1m12 was that of a SMW outflow.
Again, the formation of the SMW outflow requires explanation.

MacFadyen et al. (2001) injected jets at a radius of $R_{\rm in}=50 \km$, but their jets were injected at
a much later time in the explosion, and are less relevant to the present paper.
In any case, they also showed that SMW jets are efficient in removing
the envelope further out.
When narrow jets are simulated (e.g., Alloy et al. 2000; Zhang et al. 2003, 2004)
no envelope ejection occurs.
Indeed, narrow jets that maintain a constant direction do not expect to expel a
non-turbulent surrounding gas (Sternberg et al. 2007).
In another set of simulations, Zhang et al. (2006) added a SMW outflow to
their simulations of relativistic jets.
The papers cited above further emphasize the need to answer the question as of
how such SMW jets that can explode CCSNe are formed.

Soker (2010; hereafter paper 1) argued that SMW jets in CCSNe, as simulated by, e.g.,
Khokhlov et al. (1999) and Couch et al. (2009), can be formed when fast ($v_j > 0.3 c$)
and light (total mass of $M_{2j} \simeq 0.01-0.05 M_\odot$) jets interact with the infalling
gas at distances of ($\sim 10^3 < r < 10^4 \km$) from the center.
In paper 1 the same mechanism that was used to explain the formation of SMW jets in
other astrophysical systems was proposed to operate in CCSNe.
This mechanism is the \emph{penetrating jet feedback mechanism}, that was applied to
the growth of the super massive black hole during galaxy formation
(Soker 2009; Soker \& Meiron 2010), and to stellar environments (Soker 2008).
When the jets penetrate through the ambient medium to large distances, they don't heat
and don't expel the gas, and accretion onto the central region might continue.
When the jets don't penetrate, but rather deposit their energy
in the inner regions, they heat and expel the gas, assuming they are energetic enough and
the hot bubbles don't suffer substantial non-adiabatic cooling.

{{{ Maeda \& Nomoto (2003) studied hypernovea explosions and injected bipolar jets at a radius of
$r_{j0}=1000 \km$ for $1 \sec$, inside either a $25 M_{\odot}$ or a $40 M_{\odot}$ star.
The jets were injected at a constant direction.
Their jets had a half opening angle of either $15^\circ$ or $45^\circ$ and a kinetic energy in the
range of $1-30 \times 10^{51} \erg$.
Their results seem to indicate that the stellar mass near the equatorial plane
(perpendicular to the jets' direction) is not expelled efficiently, and the observed explosion is expected
to be highly non-spherical.
The jets we assume to be launched by the newly formed neutron star are similar
in properties to the jets simulated by Maeda \& Nomoto (2003). However, there is a substantial difference.
We assume the jets' axis to constantly vary, such that the envelope is expelled in all directions. }}}

For the efficient operation of the penetrating jet feedback mechanism in expelling the ambient gas,
the relative direction between the narrow jets and the ambient gas must be changing, such that
the jets do not manage to drill a hole and escape with their energy.
 The jets are rather shocked inside the ambient gas and heat it and/or expel it.
The change in direction between the jets and the ambient gas is achieved either by
transverse motion of the ambient gas, or changes in the jets' launching direction,
e.g., an ordered precession (Sternberg \& Soker 2008) or random direction changes termed \emph{jittering}.
{{{  Yet another way to prevent penetration is by wide jets (Sternberg et al. 2007).
In that respect, the formation of wide jets by the newly formed NS or black hole, as found by
Dessart et al. (2008), favors the operation of the model proposed here.
In the magnetic field model of Dessart et al. (2008) the angular momentum is large, and the jets are expected
to maintain a constant direction, and might have difficulties to expel the gas perpendicular to their direction.
Adding a little jittering (see section \ref{sec:ang}) will enable the jets to explode the entire envelope.
The model studied by Dessart et al. (2008) requires high initial angular momentum.
In the model discussed here there is no such requirement. }}}

Winds blown by the accretion disk around the newly formed NS have been also considered in
the literature.
MacFadyen \& Woosley  (1999) found a disk-wind in their simulations,
that has the same properties as SMW jets.
However, the disk wind blown from an extended disk surface
is less efficient (energetically speaking) than SMW jets formed from
shocked narrow jets that are launched from the very inner region of the accretion disk.
The reason is that a wind from an extended disk region is launched from
a shallow potential well.
In the penetrating jet feedback mechanism the explosion is powered by narrow jets
that are launched from the inner disk, $r \simeq 30 \km$, where the potential well is
much deeper than at the rest of the disk, and the accreted gas releases much more gravitational energy.
Indeed, MacFadyen \& Woosley  (1999) simulated the formation of a black hole (and not a NS), and
the strong disk wind is formed only when the collapsing core is rapidly rotating.

Kohri et al. (2005) conducted a study of disk wind in CCSNe, where the central object is a NS.
They propose that the wind energy is able to revive a stalled shock and
help to produce a successful supernova explosion.
Here again, the wind comes from an extended region in the disk, and it is less efficient
than fast jets blown from the very inner region of the disk.
Moreover, to form their proposed disk wind, Kohri et al. (2005) required the progenitor's
core to rotate very rapidly, as they form the accretion disk earlier than in the present model.
Another difference is that in our proposed model the jets deposit their energy further out,
and we let the material near the stalled shock to be accreted.
This is a major difference between our model and models that are based on neutrinos.
Namely, that our model does not revive the stalled shock. To the contrary, we need the material
there to be accreted and form the accretion disk that launches the jets.

In paper 1 a transverse motion of the infalling gas was considered, and a simple cooling
by neutrino formula was applied. In the present paper we consider jittering jets, and
show that for a wide range of parameters these jets can deposit their energy at
$r \sim 10^3 \km$ and form hot bubbles.
These high-pressure bubbles accelerate the rest of the star and explode it.
Before the bubbles merge they accelerate the material in their vicinity and form
the SMW outflow.
As the proposed mechanism is different in some key issues from previous models that are
based on jets and winds from the NS vicinity, we present the basic ingredients of the model
{{{  in sections \ref{sec:jet}, and \ref{sec:ang}.}}}
In section \ref{sec:pen} we use the penetrating condition to find the radius where the jets
deposit most of their energy. Neutrino cooling is discussed in section \ref{sec:cooling},
and our summary is in section \ref{sec:summary}.

\section{THE PENETRATING JET FEEDBACK MECHANISM}
\label{sec:jet}

The basic ingredients and assumptions of the proposed explosion mechanism were described in paper 1.
Here we concentrate on jittering jets, and use the following, somewhat different, assumptions and processes.
\begin{enumerate}
\item We assume that about a Chandrasekhar mass of the iron core $\sim 1.2 M_\odot$ has already collapsed,
and most of it already formed the almost final neutron star (NS). Namely, a mass of
$\sim 1.1 M_\odot$ and a radius of $\sim 20-30 \km$ is accreting the rest of the inflowing mass.
The corresponding escape velocity is $v_{\rm esc} \simeq 10^5 \km \s^{-1}$.
\item Based on recent numerical simulations (Ott et al. 2009; Blondin \& Mezzacappa 2007;
{{{  Nordhaus et al. 2010; }}} Rantsiou et al. 2010)
we assume that the angular momentum of the accreted gas onto the NS changes stochastically, and with a large
enough amplitude to form an intermittent accretion disk with a rapidly varying angular momentum direction.
Based on their simulations, we find that during a large fraction of the time the specific
angular momentum allows for the formation of an accretion disk at a radius of $\sim 30 \km$,
and that the angular momentum direction is changing on a typical rate of $\dot \phi \simeq 10 ~{\rm rad} \s^{-1}$.
In addition to the stochastic behavior behind the stalled shock at $\sim 300 \km$
(Ott et al. 2009; Blondin \& Mezzacappa 2007; Rantsiou et al. 2010),
the magnetic axes of many NS are known to be inclined to the spin.
Over all, the jets' direction is not expected to be constant, unless the core of the
collapsing star has a large amount of angular momentum.
\item The properties of jets launched by NS (or BH) have some universal properties,
such as the fast jets' speed is $v_f \simeq v_{\rm esc} \simeq 10^5 \km \s^{-1}$.
\item There is a universal ratio between mass lose rate in the two jets to mass accretion rate.
Using the same ratio of ejected to accreted energy as in Soker (2009), $0.05$, the
ejected (into both jets combined) to accreted mass average ratio is taken
to be  $\eta \equiv \dot M_f/\dot M_{\rm acc} \simeq 0.1$.
This average ratio includes the time periods when there is no accretion disk, or the disk does not blow jets.
\item The ambient density profile in the specific calculations presented here are taken to be a power law,
based on existing calculations (Wilson et al. 1986; Mikami et al. 2008)
of a $15 M_\odot$ core collapse star, rather than from the assumption of an inflow used by Soker (2010).
The mass inside that radius of $ r \la 300-1500 \km$ that is accreted and form the accretion
disk after bounce amounts to $\sim 0.1-0.5 M_\odot$.
\end{enumerate}

{From} these assumptions we derive the typical kinetic energy carried by the jets
\begin{equation}
E_j =3 \times 10^{51} \left(\frac{\eta}{0.1}\right)
\left( \frac{M_{\rm acc}}{0.3 M_\odot \s^{-1}}\right)
\left(\frac{v_f}{10^5 \km \s^{-1}}\right)^2  \erg.
\label{eq:jesenergy}
\end{equation}
With an efficiency of $k_b \sim 50 \%$ to channel the jets' energy to explosion,
these jets can account for the kinetic energy of exploding CCSNe.

\begin{figure}
\vskip -0.4cm
\includegraphics[scale=0.6]{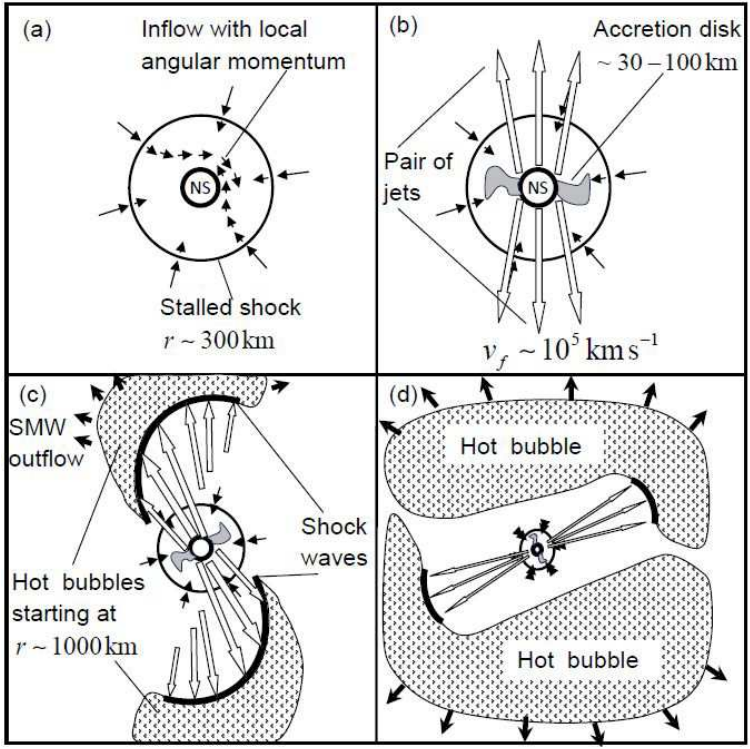}
\caption{A schematic presentation of the proposed explosion model starting after bounce (not to scale).
The four panels span a time of about one second.
(a) As a result of the stationary accretion shock instability (SASI; e.g. Blondin \& Mezzacappa 2007;
{{{  but see Nordhaus et al. 2010 for suppression of this instability in 3D simulations) }}}
or some other stochastic processes, segments of the post-shock accreted gas (inward to the stalled shock wave)
possess local angular momentum.
(b) The accreted angular momentum changes stochastically in magnitude and direction.
Namely, an intermittent accretion disk is formed or an accretion disk with a varying direction.
The disk direction, and hence the direction of the jets, changes on a rate of
$\dot \phi \simeq 10~ {\rm rad} \s^{-1}$.
(c) The jittering jets penetrate through the gas close to the center, but cannot penetrate beyond a typical
distance of $r_s \simeq 500-3000 \km$ (eq. \ref{eq:rs1}).
As the jet's axis changes its direction, no fresh jet's material
is supplied to push the jet's head forward, as depicted by the short arrows detached from the center.
The shocked jets form two hot bubbles. The bubbles accelerate the material around them
in a wide angle (shown as short-wide arrows in the upper bubble only), forming slow massive wide (SMW) jets.
This panel shows a larger volume than panel b.
(d) The inflation of several pairs of bubbles eventually eject the entire region outside
$r_s \simeq 500-3000 \km$. This explodes the star.
Unlike many other models, in the proposed model the stalled shock is not revived, but rather
the gas within $r \la 500-2000 \km$, that amounts to $\sim 0.1-0.5 M_\odot$, is accreted during the
entire jet-launching phase.
This panel shows a larger volume than panel c.}
\label{fig:jets1}
 \end{figure}

If the jets penetrate through the surrounding gas they will be collimated by that gas,
and two narrow collimated fast jets will be formed, similar to the flow structure in
the simulations of Sutherland \& Bicknell (2007) for AGN jets, and
of MacFadyen et al. (2001).
If, on the other hand, the jets cannot penetrate the surrounding gas they will deposit
their energy in the inner region.
A pair of two hot bubbles is formed. The bubbles will merge with each other and with
earlier and later pairs of bubbles.
Each bubble will accelerate the material around it over a very large solid angle.
The wide outflow has more mass than the originally narrow jets, and by
energy conservation the wide outflow has a lower velocity.
A slow massive wide (SMW) outflow (jets) has been formed.
It is this SMW outflow that expels the rest of the stellar gas.
That a wide outflow can expel the stellar gas has been studied analytically by
Kohri et al. (2005), and was demonstrated in numerical simulations
by {{{ Maeda \& Nomoto (2003) and }}}Couch et al. (2009; where their high thermal energy jets are practically SMW jets).
The results here offer an explanation to the formation of the jets simulated by
Couch et al. (2009) in their v1m12 model.
In Figure \ref{fig:jets1} we schematically summarize the proposed model.

The duration of the jets launching phase from accretion disks around super-massive black holes in AGN
is much longer than the dynamical time in the vicinity of the black hole.
This is not the situation here.
In our core collapse scenario the total accretion phase lasts for $\sim 1 \s$,
which is the typical time for the collapse of material from thousands of km.
The intermittent accretion disk changes its axis direction on a time scale of
$\sim 0.1 \s$.
The Keplerian period at the surface of the NS is $\sim 0.003 \s$, such that
the ratio of the typical time variation of the disk to the Keplerian period is only $\sim 40$.
This implies that the accretion disk might not have time to completely relax and lose its internal energy and form
a thin accretion disk.
This non-relaxed behavior might further contribute to variations in the direction of the jets,
and might increase the fraction of the accretion energy that is channeled to the outflow
(jets) kinetic energy rather than to radiation.

{{{ 
\section{ANGULAR MOMENTUM CONSIDERATIONS}
\label{sec:ang}
In our model we assume that an intermittent accretion disc residing in the region $20 \km \la r \la 50 \km$ is
formed at the final stages of the collapse. There are two limiting cases we now discuss.

In the first limit, the total angular momentum of the accreted mass is very low, i.e., the average specific angular momentum
is much lower than that of a Keplerian orbit about the radius of the NS.
In that case, we assume, based on Ott et al. (2009), Blondin \& Mezzacappa (2007), and Rantsiou et al. (2010)
that the angular momentum is rapidly changing in magnitude and direction.
As discussed above the value of the stochastically varying angular momentum amplitude can be large enough to
form an accretion disk for a very short time of $\sim 10-100 \ms$.
The direction of the angular momentum stochastically changes as well.
This stochastically variation of the angular momentum is not an independent `random walk' process.
The reason is that the total angular momentum cannot be larger than that of the initial angular momentum of
the core.
We even expect the accreted angular momentum to be {\it smaller} than the initial angular momentum of the accreted mass.
The reason is that the jets that are formed in each disk episode carry some angular momentum, and even a substantial fraction of the
angular momentum in some models (e.g., Ferreira et al. 2006).
The conclusion is that the NS will spin slower than what would be expected if the accreted mass would retain its angular momentum.
We note that the small amount of accreted angular momentum in our model is along the results of Rantsiou et al. (2010) that
found that the instability at shock does not lead to a rapidly spinning NS.
We require the typical amplitude of the variations in the specific angular momentum to be large, but expect the
total accreted angular momentum to be lower than the initial one.
Our model, therefore, fits the finding that NS are born as slow rotators (e.g., Faucher-Gigu{\`e}re et al. 2006).

The sum of the accreted angular momentum to almost zero, crudely requires the direction of the jets to rotate over $2 \pi$ during the
accretion process of $\sim 1 \s$. This is compatible with, and one of the reasons for, taking $\dot \phi \simeq 10 ~{\rm rad} \s^{-1}$.

Because of the stochastic accretion process the disk is not completely relaxed, and relative motion within the disk can
increase the viscosity.
For a disk viscosity parameter $\alpha=1$ the viscous timescale is (e.g., Dubus et al. 2001)
\begin{equation}
t_{\rm{visc}} \simeq \frac{R^2}{\nu} \simeq 10
\left(\frac{H/R_a}{0.2}\right)^{-1}
\left(\frac{C_s/v_\phi}{0.2}\right)^{-1}
\left(\frac{R}{30 \km} \right)^{3/2}
\left(\frac{M_{\rm NS}}{1.4 {M_{\odot}}}\right)^{-1/2} \alpha^{-1} \ms 
\label{eq:tvisc1}
\end{equation}
where $\nu$ is the viscosity of the disk, $H$ is the thickness of the disk,
$C_s$ is the sound speed, $v_\phi$ is the Keplerian velocity, and $M_{\rm NS}$ is the mass of the newly formed
NS or black hole.
As discussed in section \ref{sec:jet} the intermittent disk is not expected to completely settled into a thin disk.
This is the reason for taking $H/R_a \sim 0.2 $ and $C_s/v_\phi \sim 0.2$ rather than a smaller value.
We conclude that the viscosity is sufficient for accretion episodes that last $\sim 10-100 \ms$.

In the other limit the initial angular momentum of the accreted mass of the progenitor core is large, such that the disk maintains
its general direction, and the direction of the jets is constant. This is probably the case in gamma ray bursts.
If wide jets that are very efficient in expelling gas (Sternberg et al. 2007)  are formed, such as in the magnetic
model of Dessart et al. (2008), then small jittering will be sufficient to explode the star and leave a NS behind.
Such a small jittering is expected from the shock instability.
If narrow jets are formed, the small jittering will not be sufficient, unless more mass is accreted and launched into the jets.
In such a case a black hole will be formed. }}}

\section{THE NON-PENETRATION CONDITION}
\label{sec:pen}

Let the fast jets from the inner disk zone have a mass outflow rate in both directions
of $\dot M_f$, a velocity $v_f$, and let the two jets cover a solid angle of
$4 \pi \delta$ (on both sides of the disk together).
The density of the outflow at radius $r$ is
\begin{equation}
\rho_f =  \frac {\dot M_f}{4 \pi \delta r^2 v_f}.
\label{eq:rhof}
\end{equation}

{From} the numerical results for a $15 M_\odot$ star at the beginning of the
collapse as presented by Wilson et al. (1986) and Mikami et al. (2008), we find that the
collapsing material density profile can be approximated by the power law
\begin{equation}
\rho_s(r) =  \rho_0 r^{-\alpha} =
 1.3 \times 10^{10} \left( \frac {r}{100 \km} \right)^{-2.7} \g \cm^{-3},
 \qquad {\rm for} \qquad  30 \la r \la 10^4 \km.
\label{eq:rhos}
\end{equation}
The mass of the infalling gas according to equation (\ref{eq:rhos})
in the range $30 < r < 300 \km$, $30 < r < 1000 \km$, and $30 < r < 2000 \km$,
is $0.19 M_\odot$, $0.35 M_\odot$, and $0.48 M_\odot$, respectively.
In our model, most of this mass is accreted to the newly formed NS star, bringing its
mass from $\sim 1.1 M_\odot$ to a typical final mass of $\sim 1.2-1.6 M_\odot$ in most cases.
This is the mass range where most NS studied by Kiziltan et al. (2010) reside.

The head of each jet proceeds at a speed $v_h$ given by the balance
of the pressure exerted by the shocked jet's material with that
of the shocked surrounding gas.
Assuming a highly supersonic motion and neglecting the inflow velocity of the inflowing gas,
this equality reads
$\rho_s v_h^2 \simeq \rho_f (v_f-v_h)^2$, which can be solved for $v_f/v_h$
\begin{equation}
\frac {v_f}{v_h}-1 \simeq \left( \frac{\rho_s}{\rho_f} \right) ^{1/2} =
\rho_s^{1/2} \left(\frac {4 \pi \delta r^2 v_f}{\dot M_f} \right)^{1/2}.
\end{equation}
For typical parameters of our problem we find
\begin{equation}
\frac {v_f}{v_h} \simeq 2.6
\left( \frac {\dot M_f}{0.03 M_\odot \s^{-1}} \right)^{-1/2}
\left( \frac {\delta}{0.01} \right)^{1/2}
\left( \frac {v_f}{10^5 \km \s^{-1}} \right)^{1/2}
\left( \frac {r}{1000 \km} \right)^{-0.35} + 1,
\label{eq:vh1}
\end{equation}
where we took a typical mass of $\eta M_{\rm acc} \simeq 0.03 M_\odot$ to be blown into the two
jets within a typical time of 1 second.
{{{   We note that the accretion rate onto the newly formed neutron star can be quite high in the first $\sim 0.3 \s$.
In a $15 M_\odot$ it can be $\dot M_{\rm acc} \simeq 0.3 M_\odot \s^{-1}$ after $0.2 \s$
(J. Nordhaus, private communication 2011). }}}
The exponent $-0.35$ in the last term comes from $1-0.5\alpha$ for $\alpha=2.7$.
The time required for the jets to cross the surrounding gas at a distance $r$ from the center is given by
\begin{eqnarray}
t_p \simeq \frac {r}{v_h} \simeq
0.04 \left( \frac{r}{1000 \km} \right)^{0.65}
\left( \frac {\dot M_f}{0.02 M_\odot \s^{-1}} \right)^{-1/2}
\left( \frac {\delta}{0.01} \right)^{1/2}
\left( \frac {v_f}{10^5 \km \s^{-1}} \right)^{-1/2} \s .
\label{eq:tp1}
\end{eqnarray}
If the inflow velocity of the ambient gas is considered, the jets' head proceeds at a somewhat
lower velocity than that given by equation (\ref{eq:vh1}).
On the other hand, the interaction of the jets with the ambient gas can further collimate
the jets, hence reducing somewhat the value of $\delta$ and increasing the jets' head speed.

As mentioned in section \ref{sec:intro}, for the efficient operation of the penetrating jet
feedback mechanism in expelling the ambient gas, the relative direction between the narrow jets
and the ambient gas must be changing, such that the jets do not manage to drill
a hole and escape with their energy.
This implies that before a jet crosses a distance $\sim r$,
its direction should be changed on an angle larger than its width.
The time that a jet crosses its width $D_j$ is $\tau_s \equiv D_j/(\dot \phi r)$.
The width of a jet at a distance $r$ from its source is $D_j=2 r \sin \alpha$,
where $\alpha$ is the half opening angle of the jet.
For a narrow jet we have $\sin \alpha \simeq \alpha \simeq (2 \delta)^{1/2}$, and
\begin{equation}
\tau_s = \frac {2r (2 \delta)^{1/2}}{\dot \phi r}=0.03
\left( \frac{\delta}{0.01} \right)^{1/2}
\left( \frac{\dot \phi}{10 \rm \s^{-1}} \right) ^{-1} \rm s.
\label{eq:taus}
\end{equation}
{From} the demand for efficient energy deposition, $\tau_s \la t_p$, we get the condition
\begin{equation}
r_s \ga 600
\left( \frac {\eta}{0.1} \right)^{0.77}
\left( \frac {v_f}{10^5 \km \s^{1/2}} \right)^{0.77}
\left( \frac{\dot \phi}{10 \s^{-1}} \right)^{-1.54}
\left( \frac{\dot M_{\rm acc}}{0.3 M_{\odot} \s^{-1}}\right)^{0.77} \km,
\label{eq:rs1}
\end{equation}
where we used assumption 4 of section \ref {sec:jet}, $\dot M_f=\eta \dot M_{\rm acc}$,
in equation (\ref{eq:rs1}).

The main conclusion from this derivation is that the jets will stop rapid expansion
(stop penetration) at a distance of $r_s \sim 10^3 \km$ (or $r_s \sim 500 - 3000$).
This is the region from where hot bubbles will be formed.
They will accelerate the gas around them to form SMW jets, and explode the star.
It is important to note that this occurs outside the stalled shock. In our model
the stalled shock is not revived, but rather the material from its vicinity continued to be accreted
and feed the accretion disk that launches the jets.
We note that the radius of energy deposition does not depend on the half opening angle of the jets $\alpha$,
but does depend quite strongly on the rate $\dot \phi$ at which the jets' axis direction is changing.

\section{THE ROLE OF NEUTRINO COOLING}
\label{sec:cooling}

We compare the neutrino cooling time with the jets' expansion time.
The dominant process is pair production $e^+ + e^- \rightarrow \nu_{e,\mu,\tau} + \bar{\nu}_{e,\mu,\tau}$.
Based on the results of Schinder et al. (1987) and Itoh et al. (1989, 1996) we approximate the
neutrino specific cooling rate by
\begin{equation}
\epsilon_\nu \simeq 5 \times 10^{24} \left( \frac{T}{10^{10} \K} \right)^9
\erg \cm^{-3} \s^{-1}.
\label{eq:cool1}
\end{equation}

We check the cooling at the radius where the jets stop penetrating and start to inflate the bubbles.
The jets stop penetrating at $r_s \simeq 10^3 \km$ (eq. \ref{eq:rs1}).
At this radius the jets' head proceeds at a speed of $v_h\simeq 3 \times 10^4 \km \s^{-1}$ (eq. \ref{eq:vh1}).
The typical time for the jets to cross the radius $r_s$ is $\sim r_s/v_h \simeq 0.03-0.04 \s$.
Each bubble is inflated by the shocked jets' material at a supersonic speed
$u_b > C_s \simeq  10^4 \km \s^{-1}$.
We therefore take the bubbles inflation time to be $t_b \simeq  r_s/v_h \simeq 0.03-0.04 \s$.
The volume of the bubbles inflated on both sides is
\begin{equation}
V_b=2 \times \frac{4\pi}{3}(t_b u_b)^3 \simeq  2.3 \times 10^{23}
\left( \frac{u_b}{10^4 \km \s^{-1}} \right)^3
\left( \frac{t_b}{0.03 \s} \right)^3\cm^3.
\label{eq:V_bubble}
\end{equation}
The bubbles occupy $\sim 0.1$ of the volume when the jets reach $\sim 10^3$.
The total energy deposited into the bubbles is
\begin{equation}
E_b \simeq 10^{50}
\left(\frac{\eta}{0.1}\right)
\left( \frac{\dot M_{\rm acc}}{0.3 M_\odot \s^{-1}}\right)
\left(\frac{v_f}{10^5 \km \s^{-1}}\right)^2
\left(\frac{t_b}{0.03 \s} \right)\erg.
\label{eq:bubbles_energy}
\end{equation}
We emphasize that this is only the energy deposited in the first $\sim 0.03 \s$. The jets are active for
$\sim 1 \s$.
The energy in the bubbles is dominated by radiation.
We can approximate the temperature by
\begin{equation}
T \simeq \left(\frac{E_b}{aV_b}\right)^{1/4} \simeq 1.5 \times 10^{10}
\left(\frac{t_b}{0.03 \s}\right)^{-1/2} \K,
\label{eq:T_bubbles}
\end{equation}
with the other parameters as used above.

{From} equation (\ref{eq:cool1}) we find that the total energy carried by neutrinos from the bubbles when the bubbles
are inflated is
\begin{equation}
E_\nu \simeq 10^{48}    
\left(\frac{t_b}{0.03 \s}\right)^{-0.5} \erg \qquad {\rm for} \qquad t_b \ga 0.01. \s .
\label{eq:cool2}
\end{equation}
Bubbles that are inflated earlier lose more energy.
For the above parameters, the ratio of energy carried by neutrinos to that deposited into the bubbles is
\begin{equation}
\frac{E_\nu}{E_b} \simeq 0.01  
\left(\frac{t_b}{0.03 \s}\right)^{-1.5} \erg \qquad {\rm for} \qquad t_b \ga 0.01 \s.
\label{eq:cool3}
\end{equation}
Neutrino cooling is only important at very early times.
The conclusion is that neutrino cooling is not a problem to the model.

\section{SUMMARY}
\label{sec:summary}

We examined a mechanism by which narrow jets launched by the newly formed neutron star (NS)
or black hole (BH) at the center of a core collapse supernova (CCSN) lead to the supernova
explosion of the star.
Specifically, the mechanism converts fast narrow jets blown by the accreting newly formed
NS, to slow massive wide (SMW) jets, similar to those simulated by, e.g., Couch et al. (2009),
as schematically shown in Fig. \ref{fig:jets1}.
The wide outflow explosively expel the rest of the stellar mass.
The basic assumption is that after a compact NS is formed, the final accretion
stage results in the formation of an accretion disk that launches two fast narrow jets,
similar in many respects to the way young stellar objects launch jets as they accrete mass
during their final growth phase.
The other basic assumptions are listed in section \ref{sec:jet}.
The basic requirement for the jets to deposit their energy into the infalling mass
and turn the inflow to an outflow, is that the jets do not penetrate through the surrounding
infalling gas to too large distances.
For CCSNe the jets should not penetrate beyond $r_s \sim 3000 \km$.
If jets do penetrate, they carry most of their energy to distances of $r \ga 10^4 \km$,
allowing too much mass to be accreted onto the NS, hence forming a BH.
When the jets' axis does not change its direction and the surrounding gas
has no transverse velocity, the jets drill their way out (Paper 1).
For not penetrating the infalling gas the jets should encounter new material before
they manage to drill their way out.
Namely, the typical time for a jet's axis to cross the jet's width (eq. \ref{eq:taus}),
or for the infalling gas to cross the jet's width by transverse motion at radius $r_s$ (paper 1),
should be shorter than the penetration time (eq. \ref{eq:tp1}) at radius $r_s$: $\tau_s \la t_p$.
In paper 1 the transverse motion of the inflowing gas was considered.
Here we have studied the more realistic case where the jets' axis rapidly changes its direction: jittering jets.
The condition $\tau_s \la t_p$ leads to equation (\ref{eq:rs1}).

This mechanism where the negative feedback is based on the jets' non-penetrating condition
is termed the \emph{penetrating jet feedback mechanism.}
It was suggested to operate in forming SMW jets (Soker 2008) and to regulate the growth
of the super massive black hole during galaxy formation (Soker 2009; Soker \& Meiron 2010).
Namely, we suggest that NSs (or BHs) at the center of CCSNe shut off their own growth
and expel the rest of the mass available for accretion by the same mechanism, via jets,
that super massive black holes shut off their own growth, as well as that of their host bulge,
in young galaxies.
Another condition for the operation of the penetrating jet feedback mechanism is that the hot bubbles
non-adiabatic cooling time is long. In the present case the non-adiabatic cooling is due to neutrinos.
The neutrino cooling was studied in section \ref{sec:cooling} and was found to be very small.
{{{  In most cases the jittering jets will lead to a spherical explosion.
However, in some cases where the angular momentum of the accreted mass is relatively high, there will be a
preferred direction for the varying jets' axis. In such cases the explosion will
be bipolar. Such cases can explain the evidenced of asymmetry in some core collapse
supernovae (e.g., Leonard et al. 2006; Wang et al. 2003). }}}

Our model has two significant differences from most models for exploding CCSNe.
($i$) In our model neutrinos play no role in exploding the star. They can play a role
in the accretion and launching of the jets.
($ii$) In our model the stalled shock at $r \sim 300 \km$ is not revived. To the contrary,
it is required that material in the vicinity of the stalled shock is accreted and forms the
intermittent accretion disk that launches the two jets.

We note that,  contrary to intuitive expectation, when the angular momentum of the collapsing
core is very large the explosion is not efficient.
This is because the jets' axis  is constant, and the jets penetrate to large distances.
{{{  Wide jets as in the simulations of Dessart et al. (2008) can make the explosion more efficient.
In most cases with large angular momentum we expect }}}  a large amount of mass to be accreted, and forms a BH. 
The two jets that are launched by the BH might form a Gamma Ray Burst (GRB).

\acknowledgements
{{{  We thank Jason Nordhaus and the anonymous referee for helpful comments. }}}
This research was supported by the Asher Fund for Space
Research at the Technion, and by the Israel Science Foundation.

\end{document}